\documentclass[aps,pra,twocolumn,showpacs,superscriptaddress,floatfix]{revtex4}

\usepackage{dcolumn}
\usepackage{graphicx}
\usepackage{epsf}
\usepackage{amsmath}
\usepackage{amssymb}
\usepackage{bm}
\usepackage{times}

\def\Za{{Z\alpha}}

\def\rms{<\!\!\!\,r^2\!\!\!>^{1/2}}
\newcolumntype{.}{D{x}{}{-1}}

\begin{document}

\title{Non-perturbative calculation of the two-loop Lamb shift in Li-like ions}

\author{V.~A.~Yerokhin}
 \affiliation{Center for Advanced Studies, St.~Petersburg State Polytechnical
University, Polytekhnicheskaya 29, St.~Petersburg 195251, Russia}
 \affiliation{Department of Physics, St.~Petersburg State
University, Oulianovskaya 1, St.~Petersburg 198504, Russia}

\author{P.~Indelicato}
\affiliation{Laboratoire Kastler-Brossel, \'Ecole Normale
Sup\'erieure et Universit\'e P. et M. Curie, Case 74, 4
pl.~Jussieu, F-75252, France}

\author{V.~M.~Shabaev}
\affiliation{Department of Physics, St.~Petersburg State
University, Oulianovskaya 1, St.~Petersburg 198504, Russia}

\begin{abstract}

A calculation valid to all orders in the nuclear-strength parameter is presented for the two-loop
Lamb shift, notably for the two-loop self-energy correction, to the $2p\,$-$2s$ transition energies
in heavy Li-like ions. The calculation removes the largest theoretical uncertainty for these
transitions and yields the first experimental identification of two-loop QED effects in the region
of the strong binding field.

\end{abstract}


\maketitle


The Coulomb field of heavy nuclei provides a unique opportunity for testing the strong-field regime
of bound-state quantum electrodynamics (QED). The most obvious candidate for such a test is an
H-like ion, whose theoretical description is the simplest one. Measurements of the $1s$ Lamb shift
in H-like uranium, the heaviest naturally occurring element, have recently achieved an accuracy of
4.6~eV \cite{gumberidze:05:prl}. This corresponds to a fractional accuracy of 1.7\% with respect to
the total $1s$ QED contribution. Such measurements yield a test of bound-state QED at the one-loop
level (i.e., to first order in the fine structure constant $\alpha$), but they are yet insensitive
to the {\em two-loop} QED corrections, which are of primary theoretical interest at present.

By contrast, measurements of the $2p$-$2s$ transition energies in heavy Li-like ions
\cite{schweppe:91,beiersdorfer:98,brandau:04,beiersdorfer:05} have lately reached a fractional
accuracy of 0.03\% with respect to the total QED contribution. This corresponds to a 10\%
sensitivity of the experimental results to the two-loop QED contribution. These measurements
provide an excellent possibility for identification of the two-loop Lamb shift and for testing the
bound-state QED up to second order in $\alpha$ in the strong-field regime.

The theoretical description of Li-like ions is complicated by the presence of the electron-electron
interaction. For heavy ions, this interaction can be successfully accounted for within the
perturbative expansion in a small parameter $1/Z$ ($Z$ is the nuclear charge number). By
calculating a few terms of this expansion, one can rigorously describe the electron correlation and
the screening of one-loop QED corrections with the accuracy sufficient for identification of the
two-loop Lamb shift. Such a project has recently been accomplished in
\cite{yerokhin:00:prl,sapirstein:01:lamb}. Based on these calculations, an ``experimental'' value
of $-0.23$~eV was inferred in \cite{beiersdorfer:05} for the $2s$ two-loop Lamb shift in Li-like
uranium. A similar determination of the two-loop Lamb shift was earlier presented in
\cite{sapirstein:01:lamb} (based on the measurement \cite{beiersdorfer:98}) for the $2p_{3/2}$-$2s$
transition energy in bismuth.

\begin{figure}[b]
\centerline{
\resizebox{0.95\columnwidth}{!}{%
  \includegraphics{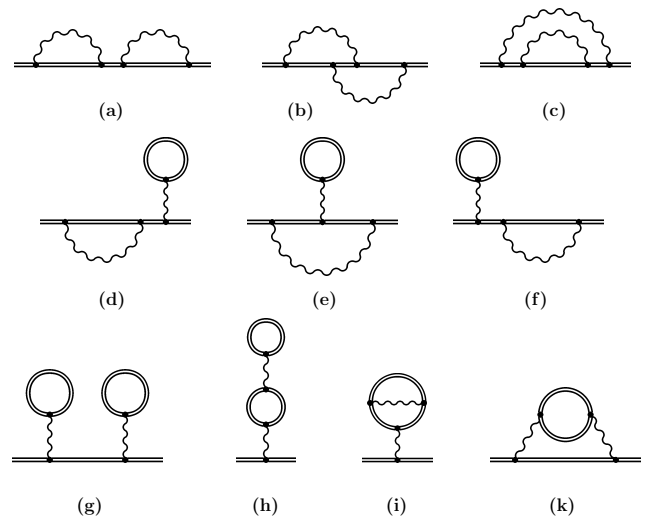}
}}
 \caption{Two-loop one-electron QED corrections. Gauge-invariant subsets are
referred to as SESE (a-c), SEVP (d-f), VPVP (g-i), S(VP)E (k).
\label{fig:2order}}
\end{figure}

The subject of our present interest is the set of two-loop one-electron QED corrections (also
referred to as the two-loop Lamb shift), graphically represented in Fig.~\ref{fig:2order}. These
corrections have lately been extensively investigated within the perturbative expansion in the
nuclear-strength parameter $\Za$
\cite{twoloop,pachucki:01:pra,jentschura:02:jpa,pachucki:03:prl,czarnecki:05:prl}. Such studies,
however, do not provide reliable information about the magnitude of the two-loop Lamb shift in
heavy ions like uranium, where the parameter $\Za$ approaches unity. Our present investigation is
addressed primarily to high-$Z$ ions and will be performed non-perturbatively, i.e., without an
expansion in $\Za$. The only exception will be made for the diagrams in
Fig.~\ref{fig:2order}(h)-(k), for which we will expand the fermion loops in terms of the binding
potential. We will keep the leading term of the expansion and refer to this as the free-loop
approximation. In the one-loop case, such approximation corresponds to the Uehling potential and
yields the dominant contribution even for high-$Z$ ions like uranium.

Necessity for a non-perturbative calculation of the two-loop Lamb shift became clear already in the
beginning of the 1990s, after the famous measurement \cite{schweppe:91} of the $2p_{1/2}$-$2s$
transition energy in U$^{89+}$ with an accuracy of 0.1~eV. Quite soon, calculations of all diagrams
in Fig.~\ref{fig:2order}(d)-(k) were accomplished \cite{beier:88} [although the graphs (i) and (k)
were calculated in the free-loop approximation only]. Three remaining diagrams (a)-(c), referred to
as the two-loop self-energy correction, turned out to be much more difficult to evaluate. The
calculation for the $1s$ state extended over a decade
\cite{mitrushenkov:95,mallampalli:98:pra,yerokhin:01:sese}, with the first complete evaluation
presented in \cite{yerokhin:03:prl} and later extended in \cite{yerokhin:05:sese}.

In this Letter, we present our calculation of the two-loop self-energy correction for the
$2s$, $2p_{1/2}$, and $2p_{3/2}$ states of several high-$Z$ ions. We also evaluate the
remaining diagrams in Fig.~\ref{fig:2order} and obtain results for the total two-loop Lamb
shift, this being previously the largest uncalculated correction for the $2p\,$-$2s$
transition energies in heavy Li-like ions. Our calculation significantly refines
theoretical predictions for these transition energies and provides a test of bound-state
QED theory in the strong-field regime up to second order in $\alpha$.

Another important aspect of our calculations is associated with their implications for the hydrogen
Lamb shift. In the previous evaluation of the $1s$ two-loop self-energy correction
\cite{yerokhin:05:sese}, we demonstrated that an extrapolation of our results to $Z=1$ yields a
value that disagrees with the analytical result to order $\alpha^2 (\Za)^6$ \cite{pachucki:03:prl}.
In view of this disagreement, it will be of interest to compare our non-perturbative values with
the analytical result to order $\alpha^2 (\Za)^6$ for the normalized difference of the $2s$ and
$1s$ energy shifts, $\Delta_s = 8\,\delta E_{2s}-\delta E_{1s}$. Such a comparison can provide us
with new information about the discrepancy for the $1s$ state, since, within the perturbative
approach, the difference $\Delta_s$ is understood much better than the energy shift of a single
$ns$ state \cite{karshenboim:94}.

The calculational scheme for the evaluation of the two-loop self-energy correction was developed
for the ground state in our previous studies \cite{yerokhin:03:prl,yerokhin:05:sese}. With this
work, we extend it to the excited states. The general procedure for isolation and cancelation of
divergences is similar to that for the $1s$ state, but the actual calculational scheme requires
substantial modifications due to a more complicated pole and angular-momentum structure of
expressions involved. Details of our calculation will be published elsewhere; in this Letter we
concentrate on presentation and analysis of the results obtained.

The two-loop self-energy correction to the energy is conveniently represented in terms of the
function $F$ defined by
\begin{equation}\label{FalphaZ}
 \delta E  = mc^2\, \left(\frac{\alpha}{\pi}\right)^2\,
                 \frac{(Z\,\alpha)^4}{n^3}\,F(Z\,\alpha)\,,
\end{equation}
where $n$ is the principal quantum number. The numerical results obtained for the $n=1$ and 2
states are listed in Table~\ref{tab:1}. The calculation was performed for the point model of the
nuclear-charge distribution.

%
%
\begin{table}[t]
\caption{The two-loop self-energy correction, in terms of
$F(\Za)$.
 \label{tab:1}}
\begin{ruledtabular}
\begin{tabular}{r....}
$Z$  &  \multicolumn{1}{c}{$1s$}
              &  \multicolumn{1}{c}{$2s$}
                             &  \multicolumn{1}{c}{$2p_{1/2}$}
                                      &  \multicolumn{1}{c}{$2p_{3/2}$}
\\  \hline\\[-9pt]
60   &  -1.66x6\, (19)  &  -1.97x6\, (70)  &    0.22x2\, (72)   &   0.02x\, (14)  \\
70   &  -1.92x3\, (18)  &  -2.45x3\, (60)  &    0.19x2\, (60)   &  -0.08x2\, (93) \\
83   &  -2.36x0\, (15)  &  -3.29x6\, (38)  &    0.13x3\, (38)   &  -0.17x5\, (66)  \\
92   &  -2.80x6\, (12)  &  -4.21x8\, (34)  &    0.01x2\, (32)   &  -0.24x1\, (48)  \\
100  &  -3.39x2\, (14)  &  -5.45x5\, (68)  &   -0.21x4\, (32)   &  -0.28x2\, (52)  \\
\end{tabular}
\end{ruledtabular}
\end{table}

As an intermediate step in our calculation, we had to consider the irreducible part of the diagram
in Fig.~\ref{fig:2order}(a), also denoted as the loop-after-loop correction, previously calculated
in \cite{mitrushenkov:95}. We report a good agreement with the previous results for the $1s$ state
but find a discrepancy for $2s$ and $2p_{1/2}$ states. For $Z=92$, we obtain $-0.090$~eV and
$-0.030$~eV for the $2s$ and the $2p_{1/2}$ state, respectively, which should be compared with the
values of $-0.069$~eV and $0.014$~eV from \cite{mitrushenkov:95}, respectively. The apparent reason
for this disagreement is a sign error in the contribution due to the imaginary part of the
self-energy operator.

The consistency of our numerical values can be tested by comparing them with analytical
results obtained within the perturbative approach. The $\Za$ expansion of the function $F$
reads
\begin{align} \label{aZexp}
 F(Z\alpha) =& \ B_{40}+
   (Z\alpha)\,B_{50} + (Z\alpha)^2\,
  \bigl[ L^3\, B_{63}
\nonumber \\ &
    +L^2\, B_{62}
 +  L\,B_{61} + G^{\rm h.o.}(\Za) \bigr]
  \,,
\end{align}
where  $L =\ln[(Z\alpha)^{-2}]$  and $G^{\rm h.o.}$ is the remainder, $G^{\rm h.o.}(\Za) = B_{60}+
\Za \,(\cdots)\,$. The coefficients $B_{40}$-$B_{61}$ are presently known for all states of our
interest \cite{twoloop,pachucki:01:pra,czarnecki:05:prl}. The coefficient $B_{60}$ was calculated
for the specific differences of energy shifts $\Delta_s = 8\,\delta E_{2s}-\delta E_{1s}$ and
$\Delta_p = \delta E_{2p_{3/2}}-\delta E_{2p_{1/2}}$ \cite{jentschura:02:jpa,czarnecki:05:prl}. A
calculation of the dominant part of $B_{60}(1s)$ and $B_{60}(2s)$ was reported in
\cite{pachucki:03:prl}, together with an estimate of unevaluated contributions to this order.

We would like now to isolate the contribution of the higher-order remainder $G^{\rm h.o.}$ from our
numerical results. Obviously, such isolation leads to a significant loss of precision, which grows
fast when $Z$ decreases. Moreover, the data presented in Table~\ref{tab:1} for the function $F$ are
already a result of a significant (and $Z$-dependent) cancelation, since individual contributions
to the energy do not exhibit the physical $Z^4$ dependence but scale typically as $Z$ or $Z^2$.
This indicates that an analysis of the $Z$ behavior of the higher-order remainder $G^{\rm h.o.}$
provides a stringent test of the consistency of the numerical results.

\begin{figure}[t]
\resizebox{\columnwidth}{!}{%
  \includegraphics{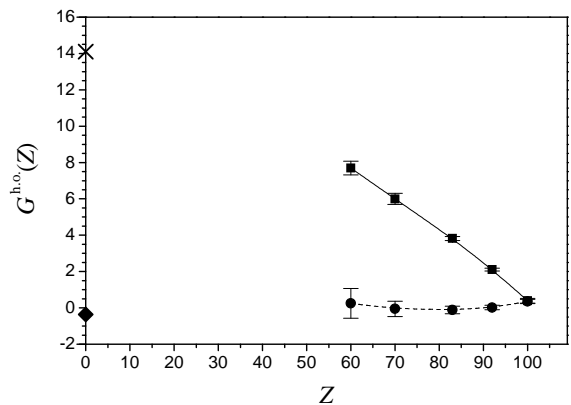}
}
 \caption{Non-perturbative results for the higher-order remainder $G^{\rm h.o.}(Z)$
 for the differences $\Delta_s$ (squares) and
 $\Delta_p$ (dots) and their limiting values at $Z=0$
denoted by the cross for $\Delta_s$ and by the diamond for $\Delta_p$.
 \label{fig:ho}}
\end{figure}

In Fig.~\ref{fig:ho}, we present our results for the higher-order remainder $G^{\rm h.o.}(Z)$ for
the differences $\Delta_s$ and $\Delta_p$, together with their limiting values $G^{\rm h.o.}(0) =
B_{60}$, which are $14.1(4)$ for $\Delta_s$ and $-0.3611$ for $\Delta_p$
\cite{jentschura:02:jpa,czarnecki:05:prl}. The comparison presented in the figure indicates that
our non-perturbative results are in good qualitative agreement with the perturbative expansion
coefficients known for the differences $\Delta_s$ and $\Delta_p$. A quantitative comparison is
presently impossible due to a lack of numerical data in the low-$Z$ region. To perform a
non-perturbative calculation for excited states of low-$Z$ ions is a difficult problem, whose
solution apparently requires development of new calculational technique. For the $1s$ state,
however, such a calculation is less problematic and we were able to carry out a direct evaluation
for $Z$ as low as 10 in our previous investigation \cite{yerokhin:05:sese}. Extrapolating our $1s$
results to $Z=1$ in that work, we found a disagreement with the analytical value for the
coefficient $B_{60}(1s)$ \cite{pachucki:03:prl}. Our present calculation for the difference
$\Delta_s$ and the agreement observed with the analytical expansion coefficients in this case can
be considered as an evidence in favor of reliability of our previous results for the $1s$ state.

In Table~\ref{tab:2}, we present the results of our calculations of all two-loop corrections
depicted in Fig.~\ref{fig:2order} for the $2p_{3/2}$-$2s$ transition in Bi$^{80+}$ and the
$2p_{1/2}$-$2s$ transition in U$^{89+}$, for which most accurate experimental data are available.
The contribution of the SESE subset is taken from Table~\ref{tab:1}. The diagrams in
Fig.~\ref{fig:2order}(d)-(g) were calculated rigorously to all orders in $\Za$, whereas the
diagrams in Fig.~\ref{fig:2order}(h)-(k) were evaluated within the free-loop approximation, i.e.,
keeping the first nonvanishing contribution in the expansion of the fermion loops in terms of the
binding potential. The error bars specified for these corrections are estimations of uncertainty
due to the approximation employed. They were obtained by multiplying the contribution of
Fig.~\ref{fig:2order}(h,i) by a factor of $(\Za)^2$ and that of Fig.~\ref{fig:2order}(k), by a
factor of $3\,(\Za)$. The factor of $3\,(\Za)$ in the latter estimation arises as a ratio of the
leading-order contribution beyond the free-loop approximation for the diagram (k), $-0.386\,
(\alpha/\pi)^2(\Za)^5$ \cite{pachucki:93:pra}, and the leading-order contribution within this
approximation, $0.142\, (\alpha/\pi)^2(\Za)^4$ \cite{lautrup:70}. The finite nuclear size effect
was taken into account in our evaluation of the diagrams in Fig.~\ref{fig:2order}(d)-(i), whereas
the other diagrams were calculated for the point nuclear model. In the case of uranium, our results
for the diagrams with closed fermion loops are in good agreement with those reported previously
\cite{mohr:98}.

%
%
\begin{table}[t]
\caption{Individual two-loop contributions to transition energies
in Li-like bismuth and uranium, in eV.
 \label{tab:2}}
\begin{ruledtabular}
\begin{tabular}{lc..}
 Subset  &  Fig.
              &  \multicolumn{1}{c}{$2p_{3/2}$-$2s$, $Z=83$}
                &  \multicolumn{1}{c}{$2p_{1/2}$-$2s$, $Z=92$}
\\ \hline\\[-9pt]
SESE     & (a)-(c)  &   0.14x5\,(4)   &     0.29x6\,(3)   \\
SEVP     & (d)-(f)  &  -0.09x5        &    -0.18x7       \\
VPVP     &  (g)     &   0.01x6        &     0.03x5       \\
VPVP     &  (h),(i) &   0.06x7\,(25)  &     0.10x1\,(46)  \\
S(VP)E   &  (k)     &  -0.01x2\,(24)  &    -0.02x2\,(45)  \\
 \hline\\[-9pt]
Total    &          &   0.12x0\,(35)  &     0.22x3\,(64)  \\
\end{tabular}
\end{ruledtabular}
\end{table}

We now turn to the experimental consequences of our calculations. In Table~\ref{tab:3}, we
collect all available theoretical contributions to the the $2p_{3/2}$-$2s$ transition
energy in Bi$^{80+}$ and to the $2p_{1/2}$-$2s$ transition energy in U$^{89+}$. The entry
labeled ``Dirac value'' represents the transition energies as obtained from the Dirac
equation with the Fermi-like nuclear potential and the nuclear-charge root-mean-square
(rms) radius fixed as $\rms = 5.851(7)$~Fm for uranium and $5.521(3)$~Fm for bismuth
\cite{angeli:04}. The dependence of the Dirac value on the nuclear model was conservatively
estimated by comparing the results obtained within the Fermi and the
homogeneously-charged-sphere models \cite{franosch:91}. We have checked that a wide class
of more general models for the nuclear-charge distribution yields results well within the
error bars obtained in this way.

The next 3 lines contain the corrections due to the one-, two-, and three-photon exchange,
respectively. QED values for the two-photon exchange correction were taken from our
previous evaluations \cite{yerokhin:00:prl,artemyev:03}. The results for the three-photon
exchange correction were obtained in this work within many-body perturbation theory (MBPT),
with retaining the Breit interaction to the first order only. For uranium, we report good
agreement with the previous evaluations of this effect \cite{zherebtsov:00}. The error
ascribed to the three-photon exchange correction is due to incompleteness of the MBPT
treatment. It was estimated by calculating the third-order MBPT contribution with two and
more Breit interactions for each state involved in the transition, adding these
contributions quadratically, and multiplying the result by a conservative factor of 2.

%
%
\begin{table}[t]
\caption{Various contributions to transition energies in Li-like bismuth and uranium, in eV.
 \label{tab:3} }
\begin{ruledtabular}
\begin{tabular}{l..}
         & \multicolumn{1}{c}{$2p_{3/2}$-$2s$,$\ Z=83$}
                              & \multicolumn{1}{c}{$2p_{1/2}$-$2s$,$\ Z=92$} \\
 \hline\\[-9pt]
Dirac value           &   2792.x21 \,(3)    &   -33.2x7 \,(9) \\
One-photon exchange   &     23.x82          &   368.8x3        \\
Two-photon exchange   &     -1.x61          &   -13.3x7        \\
Three-photon exchange &     -0.x02 \,(2)    &     0.1x5 \,(7)  \\
One-loop QED          &    -27.x48          &   -42.9x3        \\
Screened QED          &      1.x15 \,(4)    &     1.1x6 \,(3)  \\
Two-loop QED          &      0.x12 \,(4)    &     0.2x2 \,(6)  \\
Recoil                &     -0.x07          &    -0.0x7        \\
Nuclear polarization  &                     &     0.0x4 \,(2)  \\
 \hline\\[-9pt]
Total theory          &   2788.x12\,(7)    &   280.7x6 \,(14) \\
Experiment            &   2788.x14\,(4)\, \mbox{\cite{beiersdorfer:98}}
                                           & 280.6x45\,(15)\,   \mbox{\cite{beiersdorfer:05}} \\
%
\end{tabular}
\end{ruledtabular}
\end{table}

The entry labeled ``One-loop QED'' represents the sum of the first-order self-energy and
vacuum-polarization corrections calculated on hydrogenic wave functions \cite{mohr:98}. The next
line (``Screened QED'') contains the results for the screened self-energy and vacuum-polarization
corrections \cite{artemyev:99}. The uncertainty ascribed to this entry is the estimation of
higher-order screening effects; it was obtained by multiplying the correction by the ratio of the
entries ``Screened QED'' and ``One-loop QED''. The entry ``Two-loop QED'' contains the results for
the two-loop Lamb shift obtained in the present investigation. The next two lines contain the
values for the relativistic recoil correction \cite{artemyev:95} and the nuclear polarization
correction \cite{nucpol}.

Table~\ref{tab:3} shows that now, after our calculation of the two-loop Lamb shift, the total
theoretical uncertainty is significantly influenced by the error of the finite-nuclear-size effect.
It should be noted that a certain concern exists in the community about the accuracy of the
theoretical description of this correction. In particular, it was pointed out
\cite{karshenboim:05:rep} that little is known about systematical effects in experimental
determination of nuclear rms radii. In the absence of detailed investigations of such effects, we
consider the errors of rms radii obtained in \cite{angeli:04} by averaging all experimental results
available (including both muonic-ions and electron-scattering data) to be presently the most
reliable estimates and employ these values in our calculations.

The comparison presented in Table~\ref{tab:3} demonstrates that our total results agree well within
the error bars specified with the experimental data for bismuth and uranium. The theoretical
accuracy is significantly better in the former case, which is the consequence of the fact that the
finite nuclear size effect is smaller and the nuclear radius is known better. Our result for the
$2p_{3/2}$-$2s$ transition in bismuth can also be compared with the value of $2787.96$~eV obtained
by Sapirstein and Cheng \cite{sapirstein:01:lamb}. The difference of 0.16~eV between the results is
mainly due to the two-loop Lamb shift contribution (0.12~eV) which is not accounted for in
\cite{sapirstein:01:lamb}.

We conclude that inclusion of the two-loop Lamb shift is necessary for adequate interpretation of
the experimental result in the case of bismuth, whereas for uranium the two-loop Lamb shift is
significantly screened by the uncertainty due to the nuclear charge distribution. Comparison of the
theoretical and experimental results for bismuth yields the first identification of the two-loop
QED effects in the region of strong binding field, which is the first step toward the test of the
strong-field regime of bound-state QED at the two-loop level.

Valuable discussions with U.~Jentschura and K.~Pachucki are gratefully acknowledged. This work was
supported by INTAS YS grant No.~03-55-1442, by the "Dynasty" foundation, and by RFBR grant
No.~04-02-17574. The computation was partly performed on the CINES and IDRIS French national
computer centers. Laboratoire Kastler Brossel is Unit{\'e} Mixte de Recherche du CNRS n$^{\circ}$
8552.


\end{document}